Учреждение образования
«Гомельский инженерный институт Министерства по чрезвычайным ситуациям Республики Беларусь»

**Об одном методе хранения информации**

Автор:
кандидат физ.-мат. наук,
Титов Олег Владимирович

2010

# Содержание

# Введение

В современном мире одними из самых актуальных задач являются задачи передачи и хранения информации.

В данном исследовании рассматривается новый метод распределенного хранения информации.

Роль и важность системы хранения определяются постоянно возрастающей ценностью информации в современном обществе, возможность доступа к данным и управления ими является необходимым условием для выполнения любых задач.

Безвозвратная потеря данных может привести к фатальным последствиям. Утраченные вычислительные ресурсы можно восстановить, а утраченные данные, при отсутствии грамотно спроектированной и внедренной системы резервирования, уже не подлежат восстановлению.

Разработка разного рода «распределенных файловых систем» в настоящее время является одной из энергично развиваемых областей информатики. Но большинство таких систем работает на основе изготовления множества полных копий данных, хранимых в разных местах, и обеспечения различных механизмов синхронизации этих данных.

Автор исследования предлагает метод хранения информации, когда информация разделяется на маленькие «кусочки» и сохраняется на различных носителях. Предложенный автором способ позволяет восстанавливать исходную информацию имея не все «кусочки», а лишь их некоторое количество.

В отличие от современных систем хранения информации, этот способ не требует полного дублирования всех данных на разных носителях. Так же мы можем хранить все «кусочки» в открытом доступе на разных носителях, так как имея лишь один «кусочек» невозможно восстановить все данные. Последний факт позволяет использовать в качестве носителей информации различные службы хранения файлов в Интернет.

Такой подход гораздо надежнее и безопаснее, чем хранить информацию на своем собственном компьютере или на удаленном сервере, который может повредиться в любой момент. Это особенно актуально в связи с тем, что количество пользовательских данных возрастает с каждым годом в геометрической прогрессии.

# Постановка задачи

Сформулируем задачу следующим образом.

**Задача.** Разбить файл на $n$ частей таким образом, чтобы его можно было восстановить имея любые $m$ частей ($1 \leq m \leq n$).

Такие задачи называются задачами разделения секрета [1, 2, 3, 4]. Существует множество методов решения таких задач, но все они требуют достаточно большого количества вычислений применительно к поставленной выше задаче.

Наша постановка задачи отличается от классической задачи разделения секрета. Отличия заключаются в следующем:

1) главная наша задача — восстановить исходный файл, имея по крайне мере $m$ частей, тогда как в классической задаче разделения секрета главная задача — невозможность восстановить секрет при наличии меньше, чем $m$ частей. Другими словами в нашей задаче не исключена возможность восстановления исходного файла при наличии меньшего, чем $m$ частей;

2) в классической задаче разделение секрета мы под секретом понимаем число, находящееся в определенном диапазоне возможных значений, а в нашей постановке задачи мы имеем ввиду произвольный набор данных представленных как один файл.

Предложенный автором метод не требует проведения вычислений, а требуются только операции разделения файла на равные (почти равные) части и их склеивание в определенном порядке в один или несколько файлов.

# Описание метода

Пусть у нас имеется исходный файл. Требуется получить набор из $n$ файлов (будем в дальнейшем называть такие файлы модулями) таких, что имея любые $m$ модулей мы можем восстановить исходный файл.

Разобьём исходный файл на: $R = \dfrac{n-m+1}{НОД(n, n-m+1)}$ элементов (под элементами будем понимать непересекающиеся части исходного файла равного (почти равного) размера). Таким образом мы получаем, что наш исходный файл состоит из R элементов, каждому из которых можно присвоить порядковый номер от 1 до $R$.

Так как исходный файл можно восстановить имея любые $m$ модулей, то каждый из $R$ элементов должен находиться в $n - m +1$ модулей. Тогда хотя бы в одном из $m$ любых модулей будет содержаться необходимый нам элемент.

Наполнять модули элементами будем следующим образом: построим таблицу, в которой $n$ строк (количество модулей) и $K = \dfrac{n-m+1}{НОД(n, n-m+1)}$ столбцов (количество элементов находящихся в каждом модуле). Каждый из $R$ элементов должен встречаться в таблице $n - m +1$ раз, при этом он должен находиться в различных строках таблицы. Можно заполнять таблицу ставя номера элементов в каждом столбце по $n - m +1$ раз, начиная с первого столбца сверху вниз и затем переходить к следующему столбцу. И так пока не заполнится вся таблица.

**Пример.** Разбить файл на 5 частей таким образом, чтобы его можно было восстановить имея любые 3 части.

Так как НОД(5, 5 – 3+1)=1, то $R = 5$ и $K = 3$. Построим таблицу, как описано выше.

| | | | |
|---|---|---|---|
| 1: | 1 | 2 | 4 |
| 2: | 1 | 3 | 4 |
| 3: | 1 | 3 | 5 |
| 4: | 2 | 3 | 5 |
| 5: | 2 | 4 | 5 |

Т.е. первый модуль состоит из элементов 1, 2, 4; второй модуль состоит из элементов 1, 3, 4 и т.д.

Понятно, что общее количество ячеек в таблице делится на $n$ и $n - m +1$, значит, таблица имеет НОК($n$, $n - m +1$) ячеек. Тогда столбцов в таблице $K = \dfrac{НОК(n, n-m+1)}{n} = \dfrac{n-m+1}{НОД(n, n-m+1)}$, а количество элементов $R = \dfrac{НОК(n, n-m+1)}{n-m+1} = \dfrac{n}{НОД(n, n-m+1)}$.

Легко получить формулу нахождения номера элемента в таблице по его координатам:

$$N(i, j) = \left[\frac{(i-1)n + j}{n - m + 1}\right] + 1 \ ^*,$$

где i – номер строки (1 ≤ i ≤ n), j – номер столбца (1 ≤ j ≤ K).

Набор из *n* модулей, из *m* любых которых можно восстановить исходный файл, будем называть схемой (*n*, *m*).

---

* [x] – наибольшее целое число, не превосходящее x.

# Подобные схемы

Пусть дана схема ($n_1$, $m_1$), построим схему ($n_2$, $m_2$), в которой $n_2 = pn_1$ и $n_2 - m_2 + 1 = p(n_1 - m_1 + 1)$, где $p$ — рациональное, положительное число, такое что $n_2$ и $m_2$ целые.

Тогда $m_2 = 1 - p + pm_1$, или $\frac{n_2}{n_1} = p$ и $\frac{m_2 - 1}{m_1 - 1} = p$. Следовательно, $\frac{n_2}{n_1} = \frac{m_2 - 1}{m_1 - 1}$ или $\frac{n_1}{m_1 - 1} = \frac{n_2}{m_2 - 1}$.

Схемы ($n_1$, $m_1$) и ($n_2$, $m_2$), для которых выполняется равенство $\frac{n_1}{m_1 - 1} = \frac{n_2}{m_2 - 1}$, будем называть подобными.

Для подобных схем ($n_1$, $m_1$) и ($n_2$, $m_2$) выполняются следующие равенства:

$$K_1 = \frac{n_1 - m_1 + 1}{НОД(n_1, n_1 - m_1 + 1)} = \frac{p(n_2 - m_2 + 1)}{НОД(pn_2, p(n_2 - m_2 + 1))} = \frac{n_2 - m_2 + 1}{НОД(n_2, n_2 - m_2 + 1)} = K_2$$

$$R_1 = \frac{n_1}{НОД(n_1, n_1 - m_1 + 1)} = \frac{pn_2}{НОД(pn_2, p(n_2 - m_2 + 1))} = \frac{n_2}{НОД(n_2, n_2 - m_2 + 1)} = R_2$$

Значит, у подобных схем используются одни и те же элементы и размеры модулей совпадают. Более того, если $n_1 < n_2$ и схемы ($n_1$, $m_1$) и ($n_2$, $m_2$) подобны, то множество модулей схемы ($n_1$, $m_1$) является подмножеством множества модулей схемы ($n_2$, $m_2$).

**Пример.** Схемы (6, 4), (4, 3) и (2, 2) подобны.

Схему ($n$, $m$) будем называть базовой, если для любой подобной схемы ($n_1$, $m_1$) имеем $n \leq n_1$. Очевидно, что схема ($n$, $m$) является базовой тогда и только тогда, когда $НОД(n, m - 1) = 1$.

## Числовые характеристики схем

Для анализа эффективности различных схем введем числовые характеристики схем.

Избыточность – отношение размеров всех $n$ модулей к размеру исходного файла. Обозначим избыточность через $Z(n,m)$.

$$Z(n,m) = \frac{n \cdot K}{R} = n - m + 1.$$

Ясно, что чем меньше избыточность схемы, тем меньше информации (сумма размеров всех модулей) придётся рассылать и сохранять.

Введём следующую числовую характеристику показывающую на сколько маленькие модули по отношению к исходному файлу. Т.е. отношение размера одного модуля к размеру исходного файла. Обозначим эту характеристику через $ml(n,m)$.

$$ml(n,m) = \frac{K}{R} = \frac{n - m + 1}{n} = 1 - \frac{m-1}{n}.$$

Надёжность – вероятность того, что исходный файл будет восстановлен. Будем считать, что вероятности получения каждого модуля равны $p$. Обозначим надёжность через $P(n,m)$.

$$P(n,m) = \sum_{i=m}^{n} P_n(i), \text{ где } P_n(i) = C_n^i p^i (1-p)^{n-i}, \text{ т.е. } P(n,m) = \sum_{i=m}^{n} C_n^i p^i (1-p)^{n-i}.$$

Рассчитаем наименьшее число модулей для восстановления исходного файла.

В каждом модуле содержится K элементов. Каждый элемент встречается в таблице $n - m + 1$ раз. Значит имея один модуль мы получаем $K(n-m+1)$ элементов из таблицы. В таблице содержится $R(n-m+1)$ элементов. Найдём отношение общего количества элементов в таблице к количеству элементов, которое мы получаем имея один модуль. Таким образом мы получим минимальное количество модулей. Обозначим это значение через $D(n,m)$.

$$D(n,m) = -\left[\frac{-R(n-m+1)}{K(n-m+1)}\right] = -\left[\frac{-R}{K}\right] = -\left[\frac{-n}{n-m+1}\right].$$

Если схемы $(n_1, m_1)$ и $(n_2, m_2)$ подобны, то выполняются следующие равенства:

$$ml(n_1, m_1) = ml(n_2, m_2),$$
$$D(n_1, m_1) = D(n_2, m_2).$$

# Приложения метода

Используя описанный выше метод мы можем решить следующие задачи.

**Задача 1.** Пусть у нас имеется *n* носителей информации равного объёма. Требуется сохранить на них как можно больше информации и в любой момент времени восстановить её, при этом мы знаем, что одновременно могут выйти из строя не более *t* носителей.

Самый простой способ – записать одну и ту же информацию на все носители. Однако, зная что одновременно могут выйти из строя не более *t* носителей мы можем использовать схему (*n*, *m*), где *m* = *n* − *t*, записывая на каждый носитель по одному модулю.

Подсчитаем на сколько процентов возрастёт объём сохраняемой информации по сравнению со способом, когда мы сохраняем на каждый носитель одну и ту же информацию.

$$\Pr_1 = \frac{R-K}{K} \times 100\% = \frac{m-1}{n-m+1} \times 100\%.$$

**Пример.** Пусть у нас имеется 3 носителя информации равного объёма. Одновременно может выйти не более одного носителя.

Используем схему (3, 3 − 1). Тогда записывая на первый носитель первый модуль, на второй — второй, а на третий — третий, мы всегда сможем восстановить исходную информацию при выходе из строя одного из носителей.

$$\Pr_1 = \frac{3-2}{2} \times 100\% = \frac{1}{2} \times 100\% = 50\%.$$

Значит, в нашем примере мы можем сохранить на 50% больше информации.

**Задача 2.** Пусть у нас есть файл, который требуется сохранить в различных файловых хранилищах в интернете. Предположим, что из *n* ресурсов, на которые мы сохраняли файл, мы всегда можем получить доступ к *m* ресурсам. Требуется в любой момент времени иметь доступ к сохранённому файлу.

Самый простой способ — сохранить исходный файл во всех *n* файловых хранилищах..

Так как нам всегда доступны любые *m* файловых хранилищ, то мы можем использовать схему (*n*, *m*) и на каждый ресурс сохранять по одному модулю.

Подсчитаем сколько трафика (количество пересылаемой информации) мы экономим используя предложенный метод.

$$\Pr_2 = \frac{R-K}{R} \times 100\% = \frac{m-1}{n} \times 100\%.$$

**Пример.** Пусть нам требуется сохранить файл в 5-ти файловых хранилищах, при этом всегда доступны любые 3 из них. Используя схему (5, 3), в каждое хранилище записываем по одному модулю. При необходимости восстановить файл, мы скачиваем из любых доступных 3–х хранилищ модули и восстанавливаем файл.

Экономия трафика составляет $\text{Pr}_2 = \dfrac{3-1}{5} \times 100\% = \dfrac{2}{5} \times 100\% = 40\%$.

## Оптимизация метода

Введем следующую нумерацию модулей.

Первый модуль так и остаётся первым, вторым модулем назовем *n – m + 2* модуль, третьим модулем назовем *2(n – m + 1) +1* и т.д. Общая формула новой нумерации: *f(i) = (n − m +1) (i − 1) +1  mod n* (для базовой схемы). Для произвольной схемы формула будет следующей:

$$f(i) = \left( (n-m+1) \times (i-1) + 1 + \left[ \frac{(i-1) \times НОД(n, m-1)}{n} \right] \right) \bmod n ,$$

где *i* =1, 2, … , n; *f(i)* — новый номер модуля.

Подобная нумерация даёт нам следующие преимущества:
1) Каждый последующий модуль всегда состоит из элементов, которых не было в предыдущем.
2) Имея *D(n, m)* любых последовательных модулей, мы всегда восстановим исходный файл.

Т.к. *R ≤ n* , то переставим элементы в модулях таким образом, чтобы первые элементы всех модулей содержали все *R* элементов. Аналогично на втором, третьем и т.д. месте каждого модуля, тогда даже если мы получим только начало, середину или конец каждого модуля, мы сможем восстановить исходный файл.

**Пример.** Преобразуем рассмотренный пример схемы (5, 3) по приведенным выше правилам.

Начальный вариант

| | | |
|---|---|---|
| 1: | 1 | 2 | 4 |
| 2: | 1 | 3 | 4 |
| 3: | 1 | 3 | 5 |
| 4: | 2 | 3 | 5 |
| 5: | 2 | 4 | 5 |

После преобразования

| | | |
|---|---|---|
| 1: | 1 | 2 | 4 |
| 2: | 2 | 3 | 5 |
| 3: | 3 | 4 | 1 |
| 4: | 4 | 5 | 2 |
| 5: | 5 | 1 | 3 |

$$D(5, 3) = -\left[ \frac{-5}{5-3+1} \right] = \left[ \frac{-5}{3} \right] = 2 .$$

Таким образом, для восстановления исходного файла нам достаточно получить любые 2 последовательных модуля.

**Литература**